\documentclass[
  aps,prb,twocolumn,showpacs,amsmath,amssymb,superscriptaddress
]{revtex4-1}

\makeatletter
\newcommand{\whencolumns}[2]{\preprintsty@sw{#1}{#2}}
\makeatother

\usepackage[english]{babel}
\usepackage{amsmath,amssymb,amsfonts} 	% Typical maths resource package
\usepackage{graphicx}
\usepackage{array,multirow}
\usepackage[utf8]{inputenc}
\usepackage{bm}
\usepackage{xcolor}
\usepackage[normalem]{ulem}
\usepackage{siunitx}
\usepackage{hyperref}
\hypersetup{colorlinks=true,linkcolor=blue,citecolor=blue,urlcolor=blue}
\usepackage{url}
\usepackage{textcomp}
\usepackage{lineno,hyperref}
\usepackage{float}
\usepackage{textcomp} %for n^o
\usepackage{comment}
\usepackage{rotating}
\usepackage[figurename=FIG.]{caption}
\usepackage{subcaption}
\usepackage{color} % only to put comments in the draft
\definecolor{red}{rgb}{1.,0.,0.}
\usepackage{soul}
\usepackage{amsmath}
\usepackage{enumitem}

 \usepackage[
 justification=raggedright]
 {caption}
\usepackage{subcaption}
\usepackage{qcircuit}
\usepackage{braket}

\makeatletter
\newcommand*{\rom}[1]{\expandafter\@slowromancap\romannumeral #1@}
\makeatother

\usepackage{siunitx}
\usepackage{chemformula}
\usepackage{xspace}
 
\newcommand\Harvard{ John A. Paulson School of Engineering and Applied Sciences, Harvard University, Cambridge, MA 02138, USA}
\newcommand\Bosch{Robert Bosch LLC Research and Technology Center, Watertown, MA 02472, USA}

\DeclareMathAlphabet\mathbfcal{OMS}{cmsy}{b}{n}

\newcommand{\bR}{\mathbf{R}}

\newcommand{\af}{\langle\mathbf{f}\rangle}
\newcommand{\dV}{\langle\bm{\partial}^2V\rangle}

\newcommand{\mr}{\mathcal{R}}
\newcommand{\ma}{\mathcal{A}}

\begin{document}

\title{Quantum theory of nonlinear phononics}

\author{Francesco Libbi}
\affiliation{\Harvard}
\email[Corresponding author. ]{libbi@g.harvard.edu}

\author{Boris Kozinsky}
\affiliation{\Harvard}
\affiliation{\Bosch}

\begin{abstract}
The recent capability to use THz pulses to control the nuclear quantum degrees of freedom in crystals has opened promising avenues for the advanced manipulation of material properties. While numerical approaches exist for studying the time evolution of the quantum nuclear density matrix, an interpretable analytical framework to explicitly analyze the influence of quantum fluctuations on nuclear dynamics remains lacking.
In this work, we present an analytical quantum theory of nonlinear phononics. This framework is a basis for deriving models of realistic materials, allowing for exact solutions of the nuclear time evolution with full consideration of quantum fluctuations.
This is accomplished by treating for all possible third- and fourth-order phonon couplings and expressing forces as analytic functions of such fluctuations.
We provide an analytic proof that, in general, a strong pulse displacing a phonon mode from equilibrium induces the quenching, or squeezing, of its quantum lattice fluctuations. This finding, which establishes a systematization of the mechanism observed in Ref.\cite{libbi2025quantumcoolingabsolutezero}, introduces a new paradigm in nonlinear phononics, harnessing this cooling effect to drive symmetry breaking in quantum paraelectric materials.
\end{abstract}

\maketitle

\section{Introduction}

The irradiation of materials with intense THz light pulses has emerged as a powerful technique for manipulating their properties and even states of matter. In the context of light coupling to lattice vibrations, this ability gave  rise to a new branch of physics known as nonlinear phononics.
By resonantly exciting infrared-active optical phonons to large amplitudes, it is possible to leverage the anharmonicities of the crystal lattice to unlock new phases that would otherwise be inaccessible through quasi-static transformations.
This approach has been successfully employed to drive transitions such as metal-insulator \cite{becker_femtosecond_1994, cavalleri_femtosecond_2001}, para-to-ferroelectric \cite{doi:10.1126/science.aaw4913, doi:10.1126/science.aaw4911}, para-to-ferromagnetic \cite{Disa2020}, and superconducting states \cite{doi:10.1126/science.1197294}. The underlying mechanisms of these dynamical transitions have been modeled using classical oscillators coupled through third-order or fourth-order anharmonic interactions, involving one or more driven IR-active modes and additional phonon modes that are indirectly modulated \cite{Disa2021}. For example, cubic interactions such as $Q_{\mathrm{driven}}^2Q_c$ or $Q_{\mathrm{driven},1}Q_{\mathrm{driven},2}Q_c$ induce quasistatic distortions that polarize the antiferromagnet $\mathrm{CoF_2}$ \cite{Disa2020} and generate optical conductivity features similar to those of a superconducting state in several cuprate families \cite{doi:10.1126/science.1197294, Mankowsky2014} above their critical temperatures.  Furthermost, the coupling $Q_{\mathrm{driven}}^2 ·Q_c^2$ can be harnessed to induce the amplification of optical phonons \cite{doi:10.1073/pnas.1809725115}. \\
In certain classes of materials, nuclear quantum fluctuations play a critical role in the dynamics, necessitating the quantum generalization of the previously mentioned classical models. This is the case, for instance, in crystals containing light atoms \cite{Errea2016, Errea2020} or in perovskites with soft phonon modes where shallow energy barriers permit tunneling of the nuclear wavefunction \cite{PhysRevMaterials.7.L030801, Ranalli2023, PhysRevResearch.4.033020}. A pioneering study \cite{fechner_quenched_2024} of the dynamics of lattice fluctuations induced by intense THz pulses demonstrated that the cubic coupling of the antiferrodistortive (AFD) mode in $\mathrm{SrTiO_3}$ (STO) with a low-frequency acoustic ($\eta$) mode $Q_{AFD}^2Q_{\eta}$, and the quartic coupling with a resonantly driven infrared (IR) mode $Q_{AFD}^2Q_{IR}^2$ lead to the quenching of lattice fluctuations associated with the AFD mode. \\
Despite this significant progress, our understanding of the nonequilibrium dynamics of nuclear quantum fluctuations and their potential to modify crystal structures under intense laser excitations remains limited. The recently introduced time-dependent self-consistent harmonic approximation (TD-SCHA)\cite{PhysRevB.103.104305,PhysRevB.107.174307} offers the first general framework for predicting the dynamics of the full lattice fluctuation matrix. This method has enabled unprecedented insights into nuclear quantum dynamics in STO supercells \cite{lavoro_sto}, successfully reproducing the phonon upconversion observed in Ref. \cite{Kozina2019}.
In its original implementation \cite{lavoro_tecnico}, TD-SCHA requires the numerical computation of ensemble-averaged forces, typically achieved using Monte Carlo integration. While highly general and applicable to any potential energy surface (PES), this approach becomes computationally demanding for large systems. Moreover, the absence of an analytic expression for the forces makes it more challenging to fully comprehend the effects of nuclear quantum fluctuations on the dynamics. \\
This work presents a new formulation of the TD-SCHA, developed through the exact analytical evaluation of ensemble averages of forces and the curvature of the potential energy surface (PES). The theory is applicable to the most general potential energy surfaces expressed as a Taylor expansion up to fourth order, providing a quantum, fully atomistic generalization of all previously proposed nonlinear phononics models \cite{Disa2021,PhysRevB.89.220301,PhysRevLett.118.054101,PhysRevB.92.214303,PhysRevLett.129.167401,fechner_quenched_2024}.
In addition to enabling efficient first-principles simulations of quantum nuclear dynamics in real materials, our formulation serves as a foundational framework for deriving minimal models that illuminate the role of quantum fluctuations in these dynamics. We demonstrate that we recover, as a special case, the model described in Ref. \cite{fechner_quenched_2024} when considering the cubic coupling with a low frequency mode and the quartic coupling with a resonantly driven IR mode. 
Furthermore, we employ this framework to derive a model that demonstrates how the sudden excitation of a phonon mode out of equilibrium results in the cooling, or quenching, of its lattice fluctuations. This cooling reshapes the system’s potential energy landscape, facilitating access to previously inaccessible metastable states of matter \cite{libbi2025quantumcoolingabsolutezero}. We propose this mechanism as a novel paradigm in nonlinear phononics for driving light-induced phase transitions.\\
The structure of this work is organized as follows. In Sec. \ref{theory} we summarize the TD-SCHA theory. In Sec. \ref{FD}, we perform the analytical calculation of the ensemble averages of forces and the curvature of the potential energy, leading to the primary theoretical framework of this study. Sec. \ref{main_equations} derives a closed-form equation describing the dynamics of the diagonal components of lattice fluctuations. In Section \ref{model_quenching}, we demonstrate that this equation reduce to the model presented in Ref. \cite{fechner_quenched_2024} when considering the coupling with a low-frequency acoustic mode and a resonantly driven IR mode.
Section \ref{correction} discusses how the second- and fourth-derivative coefficients can be adjusted to precisely reproduce the correct shape of a double-well PES, a critical feature for accurately describing the thermodynamics and time evolution of quantum paraelectric perovskites.
Section \ref{cooling} proposes the novel paradigm for nonlinear phononics, leveraging the quenching of lattice fluctuations of a specific phonon mode to induce symmetry breaking.

\section{Theory}\label{theory}
The TD-SCHA framework \cite{PhysRevB.103.104305} models the quantum density matrix of a crystal as a Gaussian distribution in Wigner phase-space \cite{PhysRevB.107.174307}. This distribution is parametrized by the centroid positions $\mr_i$ (average atomic coordinates), centroid momenta $\mathcal{P}_i$, and the covariance matrices representing position-position, momentum-momentum, and position-momentum fluctuations, $\ma_{ij} = \braket{\delta R_i \delta R_j}$, $\mathcal{B}_{ij} = \braket{\delta P_i \delta P_j}$, and $\Gamma_{ij} = \braket{\delta R_i \delta P_j}$, respectively. All these quantities correspond to quantum fluctuations, with $\mathcal{A}_{ij}$ specifically representing the lattice fluctuations.
In cases where the interaction potential depends solely on the atomic positions, the momentum degrees of freedom can be analytically integrated out, yielding the following reduced spatial density
\begin{equation}\label{rho}
    \rho(\bR,t) = \frac{1}{\mathcal{N}}e^{-\frac{1}{2} \sum_{ij}(R_{i}- \mr_i ) \ma^{-1}_{ij}(t)(R_{j} - \mr_j) } \ ,
\end{equation}
where $\mathcal{N}$ is the normalization factor. 
All variables are rescaled by the atomic masses, as outlined in Ref. \cite{PhysRevB.107.174307} and summarized in Appendix \ref{scaling_rules} (e.g., \(\mr_i = \tilde\mr_i \sqrt{m_i}\), with \(\tilde\mr_i\) denoting the unscaled position). The indices $i$ and $j$ enumerate both atomic sites in the supercell and Cartesian coordinates. 
The expectation value of an operator that is a function of the atomic positions is given by:
\begin{equation}
    \braket{O}(t) = \int d\bR O(\bR)\rho(\bR,t) \ .
\end{equation}
The nuclear density in Eq. \ref{rho} is evolved by solving the following differential equations:
\begin{equation}\label{tdscha}
    \begin{cases}
        \dot{\mr}_i = \mathcal{P}_i \\
        \dot{\mathcal{P}_i} = \braket{f}_i \\
        \dot{\ma}_{ij} = \Gamma_{ij} + \Gamma_{ji} \\
        \dot{\mathcal{B}}_{ij} = -\sum_{\kappa}\braket{\partial^2V}_{ik}\Gamma_{kj} - \sum_{\kappa}\braket{\partial^2V}_{jk}\Gamma_{ki}\\
        \dot{\Gamma}_{ij} = \mathcal{B}_{ij} - \sum_{\kappa} \ma_{ik}\braket{\partial^2V}_{kj}
    \end{cases}\ ,
\end{equation}
which are obtained starting by the interacting phonon Hamiltonian and imposing the least action principle \cite{PhysRevB.107.174307}. Even though the matrices $\mathcal{B}$ and $\Gamma$ have been integrated out from the density in Eq. \ref{rho}, their evolution can not be disentangled from that of the other parameters (unless approximations are adopted, as shown in Sec. \ref{main_equations}).
The atomic potential energy landscape
\begin{equation}
    V(\bR, t) = V_\text{BO}(\bR) + V_\text{ext}(\bR, t).
\end{equation}
determines the quantum averages of the forces, $\af$, and the average curvature tensor, $\dV$, defined as:
\begin{equation}
    \langle f_{i}\rangle(t) = -\int d\bR \frac{\partial V}{\partial R_{i}}(t) \rho(\bR, t),
    \label{eq:force}
\end{equation}
\begin{equation}
    \langle\partial_{ij}^2V\rangle(t) = \int d\bR \frac{\partial^2 V}{\partial R_{i} \partial R_{j}}(t)\rho(\bR, t)\ .
    \label{eq:d2v}
\end{equation}
Here $V_\text{BO}(\bR)$ is the instantaneous interaction potential of nuclei within the Born-Oppenheimer approximation that depends only on the nuclear positions, and $V_\text{ext}(\bR, t)$ is an external time-dependent potential. In this work, we assume that  $V_\text{ext}$  is a linear function of the atomic positions. This assumption is valid, for instance, in the case of dipole coupling with an electric field. Under this condition,  $V_\text{ext}$  does not contribute to the ensemble average of the curvature, as expressed in Eq. \ref{eq:d2v}. \\
Eqs. \ref{eq:force} and \ref{eq:d2v} involve multidimensional integrals, whose evaluation is highly challenging. An efficient and versatile approach to compute these integrals is through Monte Carlo sampling \cite{lavoro_tecnico}. To eliminate the effect of stochastic noise, which could render the equations unstable, the set of random configurations should be generated at the start of the evolution and subsequently evolved consistently with the nuclear distribution \cite{lavoro_tecnico}. This approach is highly general and applies to any type of interaction potential. However, when the potential energy landscape can be accurately modeled using a fourth-order Taylor expansion—as is the case for many systems of interest for nonlinear phononic phenomena—the ensemble averages in Eqs. \ref{eq:force} and \ref{eq:d2v} can be computed analytically. This enables the formulation of a comprehensive quantum theory for nonlinear phononics, where the influence of lattice fluctuations on the dynamics is explicitly captured in analytical expressions. Moreover, the exact computation of ensemble averages significantly reduces computational cost, facilitating the use of high-order integration schemes. In the next section, we calculate the ensemble average of the potential energy and its derivatives, resulting in a reformulation of the TD-SCHA.
 
\section{Exact analytic averages}\label{FD}
The nonlinear phononics phenomena studied in recent years have been modeled using anharmonic processes involving third- or fourth-order phonon couplings. To provide the most comprehensive treatment, we assume that the interaction potential can be expressed as a general fourth-order polynomial. For clarity and conciseness, we adopt the Einstein summation convention for repeated indices throughout this section:
\begin{equation}\label{Vu}
	V(u)  = \frac{1}{2}\phi_{ij}u_iu_j + \frac{1}{3!}\chi_{ijk}u_iu_ju_k + \frac{1}{4!}\psi_{ijkl}u_iu_ju_ku_l\ ,
\end{equation}
where $u_i$ represents the displacement of the i-th atom relative to the reference structure, about which the Taylor expansion of the potential is calculated.  To minimize the number of terms required to describe the potential energy surface, it is convenient to use the most symmetric structure compatible with the geometry of the chosen unit cell as the reference.\\
The ensemble average of a polynomial potential over the Gaussian distribution in Eq. \ref{tdscha} can be computed analytically, expressing $u_i=\mr_i+ \delta R_i$, where $\delta R_i = R_i - \mr_i$, and using the following relations:
\begin{equation}\label{RR}
	\braket{\delta R_i \delta R_j} = \ma_{ij},
\end{equation}
and
\begin{equation}\label{RRRR}
	\braket{\delta R_i \delta R_j \delta R_k \delta R_l} = \ma_{ij}\ma_{kl} + \ma_{ik}\ma_{jl} + \ma_{il}\ma_{jk}\ ,
\end{equation}
whereas the product of odd terms averages to zero:
\begin{equation}
    \braket{\delta R_i} = \braket{\delta R_i\delta R_j\delta R_k} = 0 \ .
\end{equation}
We emphasize that, in this formalism, the centroid coordinates are expressed as relative displacements with respect to the Cartesian coordinates of the reference structure used for the Taylor expansion.
The ensemble average of the potential reads 
\begin{equation}
\begin{aligned}
        \braket{V} = V(\mr) + \frac{1}{2}\phi_{ij}\ma_{ij} + \frac{1}{2}\chi_{ijk}\mr_i\ma_{jk} \\ + \frac{1}{4}\psi_{ijkl}\mr_i\mr_j\ma_{kl} 
        +\frac{1}{8}\psi_{ijkl}A_{ij}A_{kl} \ .
\end{aligned}
\end{equation}
Here $V(\mr)$ is the same anharmonic potential expressed in Eq. \ref{Vu}, as a function of the centroid coordinates. 
Using the same rules, it is possible to compute the ensemble average of the forces as:
\begin{equation} \label{force}
	\begin{aligned}
		\braket{f_i} = f_i(\mr) - \frac{1}{2}\chi_{ijk}\ma_{jk} - \frac{1}{2}\psi_{ijkl}\mr_j\ma_{kl},
	\end{aligned}
\end{equation}
where
\begin{equation}
	f_i(\mr) = - \phi_{ij}\mr_j - \frac{1}{2}\chi_{ijk}\mr_j\mr_k - \frac{1}{3!}\psi_{ijkl}\mr_j\mr_k\mr_l
\end{equation}
represents the classical force acting on the system.
From Eq. \ref{force}, it is evident that the quantum force is given by:
\begin{equation}
	f_i^q(\mr) = - \frac{1}{2}\chi_{ijk}\ma_{jk} - \frac{1}{2}\psi_{ijkl}\mr_j\ma_{kl}.
\end{equation}
We observe that the quantum forces arising from third-order interactions couple directly with the lattice fluctuations matrix $\ma_{jk}$, while those originating from fourth-order interactions result from a hybrid coupling between fluctuations and the centroid displacement. Notably, the quantum force resulting from fourth-order interactions vanishes for the reference configuration ($\mr_i = 0$). \\
Finally, the ensemble average of the curvature reads
\begin{equation}\label{curv}
   \braket{\partial^2V_{ij}} = \phi_{ij} + \chi_{ijk}\mr_k + \frac{1}{2}\psi_{ijkl}\mr_k\mr_l  + \frac{1}{2}\psi_{ijkl}\ma_{kl}\ .
\end{equation}
The quantum correction to the curvature, represented by the term  $\frac{1}{2}\psi_{ijkl}\ma_{kl}$, has the potential to stabilize phonons that are inherently unstable at the harmonic approximation.\\
The differential equations in Eq. \ref{tdscha}, together with Eqs. \ref{force} and \ref{curv}, form a closed problem that can be integrated to study the quantum dynamics of atoms.\\
The initial condition for Eqs. \ref{tdscha} corresponds to thermodynamic equilibrium, with parameters determined using the Self-Consistent Harmonic Approximation (SCHA) \cite{PhysRevB.96.014111, Monacelli_2021}. This method minimizes the free energy functional:
\begin{equation}\label{free}
    \mathcal{F} = \braket{K+V}-T\braket{S}\ .
\end{equation}
Here $K$ is the kynetic energy, $T$ is the temperature and $S$ is the entropy. 
The minimization of the free energy in Eq. \ref{free} is equivalent to solving the following equations \cite{PhysRevB.96.014111, Monacelli_2021}:
\begin{equation}\label{sscha}
\begin{cases}
    \braket{f}_{(\mr,\ma[\Phi])}=0 \\
    \Bigl\langle \partial^2 V\Bigr\rangle_{(\mr,\ma[\Phi])} = \Phi\ ,
\end{cases}
\end{equation}
where $\Phi$, the SCHA dynamical matrix, is obtained self-consistently. The initial values for the lattice fluctuations are
\begin{equation}\label{A_eq}
    \ma_{ij}[\Phi] = \sum_{\mu}\frac{\hbar(2n_{\mu} +1)}{2\omega_{\mu}} e_{\mu i }e_{\mu j }\ , 
\end{equation}
\begin{equation}\label{B_eq}
	\mathcal{B}_{ij}[\Phi] = \sum_{\mu}\frac{\hbar(2n_{\mu} +1)\omega_{\mu} }{2} e_{\mu i }e_{\mu j }\ , 
\end{equation}
and
\begin{equation}\label{eq_gamma}
	\Gamma_{ij} = 0\ ,
\end{equation}
where $n_{\mu}$ is the Bose-Einstein distribution, and $\omega_{\mu}^2$ and $e_{\mu}$ are the eigenvalues and eigenvectors of the matrix $\Phi$, respectively. These represent the auxiliary anharmonic phonons \cite{Monacelli2021} and will serve as a reference in the theory developed later. Notably, the conditions in Eqs. \ref{sscha}-\ref{eq_gamma}  correspond to the stationary points of Eqs. \ref{tdscha}, highlighting the compatibility between the two theories. Specifically, substituting Eqs. \ref{sscha}-\ref{eq_gamma} into the last of Eqs. \ref{tdscha} yields:
\begin{equation}\label{equilibrium}
	\dot{\ma}_{ij} = \dot{\mathcal{B}} _{ij}= \dot{\Gamma}_{ij} = 0 \ ,
\end{equation}
indicating that the quantum fluctuations remain constant over time at the thermodynamic equilibrium. 

\section{Disentangled equation for lattice fluctuations}\label{main_equations}
In this section, we derive an equation that relates the evolution of the diagonal components of the lattice fluctuations to the Cartesian degrees of freedom alone, thereby decoupling it from the dynamics of $\mathcal{B}$ and $\Gamma$.
We start by projecting Eq. \ref{tdscha} onto the phonon eigenvectors $e_{\mu i}$ defined in Sec. \ref{FD}, where $\mu$ denotes the phonon branch. 
The projection of a generic tensor $T_{ij\ldots k}$ is achieved by contracting its indices with the Cartesian indices of the phonon eigenvectors, $T_{\mu\nu...\sigma} = \sum_{ij...k}T_{ij...k}\ e_{\mu i}e_{\nu j}...e_{\sigma k}$. Throughout this work, we adopt the convention that tensors with Greek indices are understood to be projected onto the phonon eigenvector basis. For example,
\begin{equation}
    \psi_{\mu\nu\sigma\tau} = \sum_{ijkl}\psi_{ijkl}\ e_{\mu i}e_{\nu j}e_{\sigma k}e_{\tau l}
\end{equation}
represents the fourth derivative computed along the phonon coordinates.
Furthermore, we will adopt the shorthand $\kappa_{\mu\nu} = \braket{\partial^2V}_{\mu\nu}$.
Projecting Eq. \ref{tdscha} onto the phonon coordinates is thus formally equivalent to substituting all Cartesian indices with Greek indices (see Eq. \ref{tdscha_mu} in Appendix \ref{eq_A}).\\
In general, disentangling the equations governing lattice fluctuations is not feasible without approximations, due to the asymmetry of the $\Gamma$ matrix and the time dependence of the curvature. However, under the assumption that the off-diagonal coupling between the curvature and the fluctuation matrices is negligible, specifically:
\begin{equation}\label{approx_a_text}
\sum_{\tau}\kappa_{\mu\tau}\ma_{\tau\mu} \approx \kappa_{\mu\mu}\ma_{\mu\mu}
\end{equation}
and
\begin{equation}\label{approx_gamma_text}
\sum_{\tau}\kappa_{\mu\tau}\Gamma_{\tau\mu} \approx \kappa_{\mu\mu}\Gamma_{\mu\mu},
\end{equation}
it becomes possible to derive a decoupled differential equation for the diagonal components $\ma_{\mu\mu}$ (see Appendix \ref{eq_A} for the detailed derivation):
\begin{equation}\label{main}
\dddot{\tilde\ma}_{\mu\mu} + 4\dot{\tilde\ma}_{\mu\mu}\kappa_{\mu\mu} + 2\ma_{\mu\mu}\dot{\kappa}_{\mu\mu} = 0\ ,
\end{equation} 
where the diagonal components of the curvature and their time derivatives are given by:
\begin{equation}\label{kmu1}
\kappa_{\mu\mu} = \omega^2_{\mu} + \sum_{\tau}\chi_{\mu\mu\tau}\tilde\mr_{\tau} + \sum_{\tau\sigma}\frac{1}{2}\psi_{\mu\mu\tau\sigma}(\tilde\mr_{\tau}^2 + \tilde A_{\tau\sigma}) \ ,
\end{equation}
\begin{equation}\label{kmu_dot1}
\dot{\kappa}_{\mu\mu} = \sum_{\tau}\chi_{\mu\mu\tau}\dot{\tilde\mr}_{\tau} + \sum_{\tau\sigma} \frac{1}{2}\psi_{\mu\mu\tau\sigma}(2\dot{\tilde\mr}_{\tau}\tilde\mr_{\sigma} + \dot{\tilde A}_{\tau\sigma}) \ .
\end{equation}
Here $\tilde\mr_\mu = \mr_{\mu} - \mr_{\mu}^{eq}$ and $\tilde\ma_{\mu\nu} = \ma_{\mu\nu} - \ma_{\mu\nu}^{eq}$ denote deviations from the equilibrium values. The initial condition for these equations are 
\begin{equation}\label{ic}
   \tilde\ma(0)=\dot{\tilde\ma}(0)= \ddot{\tilde\ma}(0)=\tilde\mr(0)=0\ .
\end{equation}
At equilibrium, Eqs. \ref{approx_a_text} and \ref{approx_gamma_text} hold exactly due to two independent factors, either of which alone would be sufficient to ensure their validity. The first factor is that the curvature $\kappa_{\mu\nu}$ is diagonal. The second is that \(\ma_{\mu\nu}\) is diagonal and $\Gamma_{\mu\nu}$ vanishes identically. As a result, these equations are expected to remain highly accurate approximations even when the system is slightly perturbed from equilibrium \cite{PhysRevB.107.174307}.
In the strongly nonequilibrium regime, Eqs. \ref{approx_a_text} and \ref{approx_gamma_text} are expected to hold as long as at least one of the factors remains approximately valid. We now provide conditions under which this is true. When a specific phonon mode $\mu$ is selectively excited, then it is reasonable to assume $|\tilde\mr_{\mu}|\gg|\tilde\mr_\nu|$ $\forall\ \nu\neq\mu$ and $|\tilde\ma_{\mu\mu}|\gg|\tilde\ma_{\sigma\tau}|$ $\forall\ (\sigma,\tau)\neq(\mu,\mu)$. The expression for the curvature will then reads
\begin{equation}
    \kappa_{\mu\nu} \simeq \omega^2_{\mu}\delta_{\mu\nu} + \chi_{\mu\nu\mu}\tilde\mr_{\mu} + \frac{1}{2}\psi_{\mu\nu\mu\mu}(\tilde\mr_{\mu}^2+ \tilde\ma_{\mu\mu})\ . 
\end{equation}
For $\kappa$ to remain approximately diagonal, the following condition must be satisfied:
\begin{equation}
    \Bigr|\omega^2_\mu + \frac{1}{2}\psi_{\mu\mu\mu\mu}(\tilde\mr_{\mu}^2 + \tilde\ma_{\mu\mu}) \Bigl| \gg \Bigl|\frac{1}{2}\psi_{\mu\nu\mu\mu}(\tilde\mr_{\mu}^2 + \tilde\ma_{\mu\mu}) 
    \Bigr| \ ,
\end{equation}
a condition that holds as long as $\psi_{\mu\mu\mu\mu}$ is greater than $\psi_{\mu\mu\mu\nu}$. Furthermore, from the differential equation governing the evolution of $\Gamma$,
\begin{equation}
    \dot{\Gamma}_{\mu\nu}=\mathcal{B}_{\mu\nu} + \sum_{\sigma}\ma_{\mu\sigma}\kappa_{\sigma\nu}\ .
\end{equation}
we can see that, at the beginning of the evolution, the diagonal components of $\Gamma$ grow much faster than the off-diagonal components, as all the matrices on the right-hand side are diagonal.
Eq. \ref{approx_a_text} and \ref{approx_gamma_text} could break down, however, in the case of resonant excitation of phonon modes in different directions, particularly in systems with low symmetry where strong off-diagonal couplings between phonon modes are expected.
In such cases, solving the full set of equations in \ref{tdscha} is necessary to accurately simulate the quantum evolution of the system.

\section{Reduction to established model}\label{model_quenching}
In this section, we demonstrate that, when considering coupling to only few selected degrees of freedom, Eqs. \ref{main}, \ref{kmu1}, and \ref{kmu_dot1} simplify to the model for lattice fluctuation dynamics \cite{Garrett:97} that has been successfully applied to describe the quenching of lattice fluctuations in the AFD mode of STO, as reported in Ref. \cite{fechner_quenched_2024}.
Let us assume that the mode $\mu$ is coupled to the mode $\eta$ through cubic interaction, and to a mode $\iota$ through quartic interaction, described by the potential:
\begin{equation}\label{almost_final}
    V = \frac{1}{3!}\chi_{\mu\mu\eta} \mr_{\mu}^2\mr_{\eta} + \frac{1}{4!} \psi_{\mu\mu\iota\iota}\mr_{\mu}^2\mr_{\iota}^2\ .
\end{equation}
For this interaction potential, the ensemble average of the curvature in Eq. \ref{kmu1} and its time derivative reduce to:
\begin{equation}\label{kmu1_model}
\kappa_{\mu\mu} = \omega^2_{\mu} + \chi_{\mu\mu\eta}\tilde\mr_{\eta} + \frac{1}{2}\psi_{\mu\mu\iota\iota}(\tilde\mr_{\iota}^2 + \tilde A_{\iota\iota}) \ ,
\end{equation}
\begin{equation}\label{kmu_dot1_mode}
\dot{\kappa}_{\mu\mu} = \chi_{\mu\mu\eta}\dot{\tilde\mr}_{\eta} + \frac{1}{2}\psi_{\mu\mu\iota\iota}(2\dot{\tilde\mr}_{\iota}\tilde\mr_{\iota} + \dot{\tilde A}_{\iota\iota}) \ .
\end{equation}
By defining $g_{\mu,\eta} = \chi_{\mu\mu\eta}$ and $g_{\mu,\iota} = \frac{1}{2}\psi_{\mu\mu\iota\iota}$, and combining Eqs. \ref{main} with \ref{kmu1_model} and \ref{kmu_dot1_mode} leads to
\begin{equation}\label{final_model}
\begin{aligned}
        \dddot{\ma}_{\mu\mu} + 4\dot{\ma}_{\mu\mu}\Bigl(\omega_{\mu}^2+g_{\mu,\eta}\tilde\mr_{\eta} + g_{\mu,\iota}\tilde\mr_{\iota}^2 + g_{\mu,\iota}\tilde\ma_{\iota\iota}\Bigr) =\\
        -2\ma_{\mu\mu}\Bigl( g_{\mu,\eta}\dot{\tilde\mr}_{\eta} +  2g_{\mu,\iota}\dot{\tilde\mr}_{\iota}\tilde\mr_{\iota} + g_{\mu\iota}\dot{\tilde\ma}_{\iota\iota}  \Bigr) \ .
\end{aligned}
\end{equation}
Setting $\mu=\mathrm{AFD}$ and $\iota=\mathrm{IR}$, and recalling that $\ma_{\mu\mu}=\braket{u_{\mu}^2(t)}$, it is very straightforward to see that this equation coincides with Eq. 3 of Ref. \cite{fechner_quenched_2024} apart from the term proportional to $g_{\mu,\iota}\tilde\ma_{\iota\iota}$ (and its time derivative), and the dissipative component $\gamma\ddot{\mathcal{A}}_{\mathrm{AFD}}$.
The contribution of the first term may be negligible when the mode $\iota$ is driven resonantly, as it may be reasonable to assume $g_{\mu,\iota}\tilde\mr_{\iota}^2 \gg g_{\mu,\iota}\tilde\ma_{\iota\iota}$. 
The second term represents energy exchange between degrees of freedom and heat transfer to an external bath. TD-SCHA captures the former through interactions with other phonon modes, which are neglected in this simplified model. Since TD-SCHA conserves the total quantum energy of the system, it does not account for external heat exchange. However, such exchange is expected to be negligible when studying ultrafast phenomena. It is important to emphasize that, while the approximations in Eqs. \ref{approx_a_text} and \ref{approx_gamma_text}, along with the neglect of the term \(g_{\mu,\iota}\tilde{\ma}_{\iota\iota}\), are reasonable under certain regimes, their validity is not universal. This highlights the importance of the formalism introduced in Eqs. \ref{tdscha}, which provides a complete and accurate description of the dynamics regardless of the specific features of the material or the laser excitation.

\section{Correction of the shape of a double-well PES}\label{correction}
The accuracy of TD-SCHA is inherently tied to the method used for calculating energy and forces. Despite significant advancements in first-principles simulations and machine learning techniques, certain materials exhibit complex features in their potential energy surfaces that remain challenging to fully capture.
A notable example is the double-well PES of the soft modes in perovskites \cite{PhysRevMaterials.7.L030801, Ranalli2023}, which is highly sensitive to the choice of the functional in DFT calculations. The challenge lies in the fact that no single functional reliably captures all relevant features. For instance, in STO, RPA calculations accurately reproduce the ferroelectric (FE) soft phonon frequency in agreement with experimental results but fail to provide a correct estimate of the AFD transition temperature \cite{PhysRevMaterials.7.L030801}.
Since both the FE and AFD modes are expected to play a critical role in the nonequilibrium physics of STO at low temperatures, it is crucial to develop an approach that accurately captures their PES shapes. Representing the potential energy as a Taylor expansion up to fourth order provides a straightforward method to tune the PES for multiple phonon modes simultaneously. A double-well PES can be described using a quartic function of the form:
\begin{equation}\label{V_1d}
    V(u) = \frac{1}{2}\phi u^2 + \frac{1}{4!}\psi u^4\ ,
\end{equation}
with $\phi < 0$. Defining the depth of the well as: 
\begin{equation}
    V^0 = \mathrm {min}_x V
\end{equation}
and the distance of the well’s minimum from the origin:
\begin{equation}
    x^0 = \mathrm{argmin}_x\ V\ ,
\end{equation}
it is possible to reparameterize the potential as a function of these variables \cite{PhysRevResearch.4.033020}:
\begin{equation}
    \frac{V(u)}{V^0} = -2\Bigl(\frac{u}{x^0}\Bigr)^2+\Bigr(\frac{u}{x^0} \Bigl)^4\ .
\end{equation}
The relationship between the coefficient $\phi$ and $\psi$ and the parameters $V_0$ and $x_0$ is given by:
\begin{equation}\label{refphi}
    \phi = -\frac{4V^0}{{x^0}^2}\ ,
\end{equation}
and
\begin{equation}\label{refpsi}
    \psi = \frac{4!V^0}{{x^0}^4}\ .
\end{equation}
To assign the desired values $V^0_{\mu}$ and $x^0_{\mu}$ for a mode $\mu$, while minimally affecting other couplings, we first project the second and fourth derivatives into the phonon space:
\begin{equation}
    \phi_{\mu\nu} = \sum_{ij}\phi_{ij} e_{\mu i}e_{\nu j} 
\end{equation}
\begin{equation}
    \psi_{\mu\nu\sigma\tau} = \sum_{ijkl}\psi_{ijkl} e_{\mu i}e_{\nu k}e_{\sigma k}e_{\tau l}\ .
\end{equation}
(Note that all lengths in this framework are mass-rescaled, such that $\phi_{ij}=\frac{\bar{\phi}_{ij}}{\sqrt{m_im_j}}$ and $\psi_{ij}=\frac{\bar{\psi}_{ijkl}}{\sqrt{m_im_jm_km_l}}$, where $\bar{\phi}$ and $\bar{\psi}$ represent the unscaled variables).
Next, the values of  $\phi_{\mu\mu}$ and $\psi_{\mu\mu\mu\mu}$ are determined based on the desired values for the height of the potential barrier $V^0_{\mu}$ and the position of the minimum $x^0_{\mu}$:
\begin{equation}
    \phi_{\mu\mu} = -\frac{4V^0_{\mu}}{{x^0_{\mu}}^2}
\end{equation}
\begin{equation}
    \psi_{\mu\mu\mu\mu} = \frac{4!V^0_{\mu}}{{x^0_{\mu}}^4}
\end{equation}
After adjusting these coefficients, the derivatives are projected back into the Cartesian basis. Due to the orthogonality of phonon modes, modifying the diagonal components of the tensors $\phi$ and $\psi$ for mode $\mu$ does not influence the  components of any other mode $\nu$. Furthermore, since the typical value of $V^0$ for perovskites is very small (on the order of a few meV per formula unit), the changes to the derivative tensors $\phi$ and $\psi$ are negligible.  Nevertheless, even slight corrections to the diagonal components of the PES for the soft modes remain critical due to the extreme sensitivity of their frequencies \cite{PhysRevMaterials.7.L030801}.
 
\begin{figure}
    \centering
    \includegraphics[width=\linewidth]{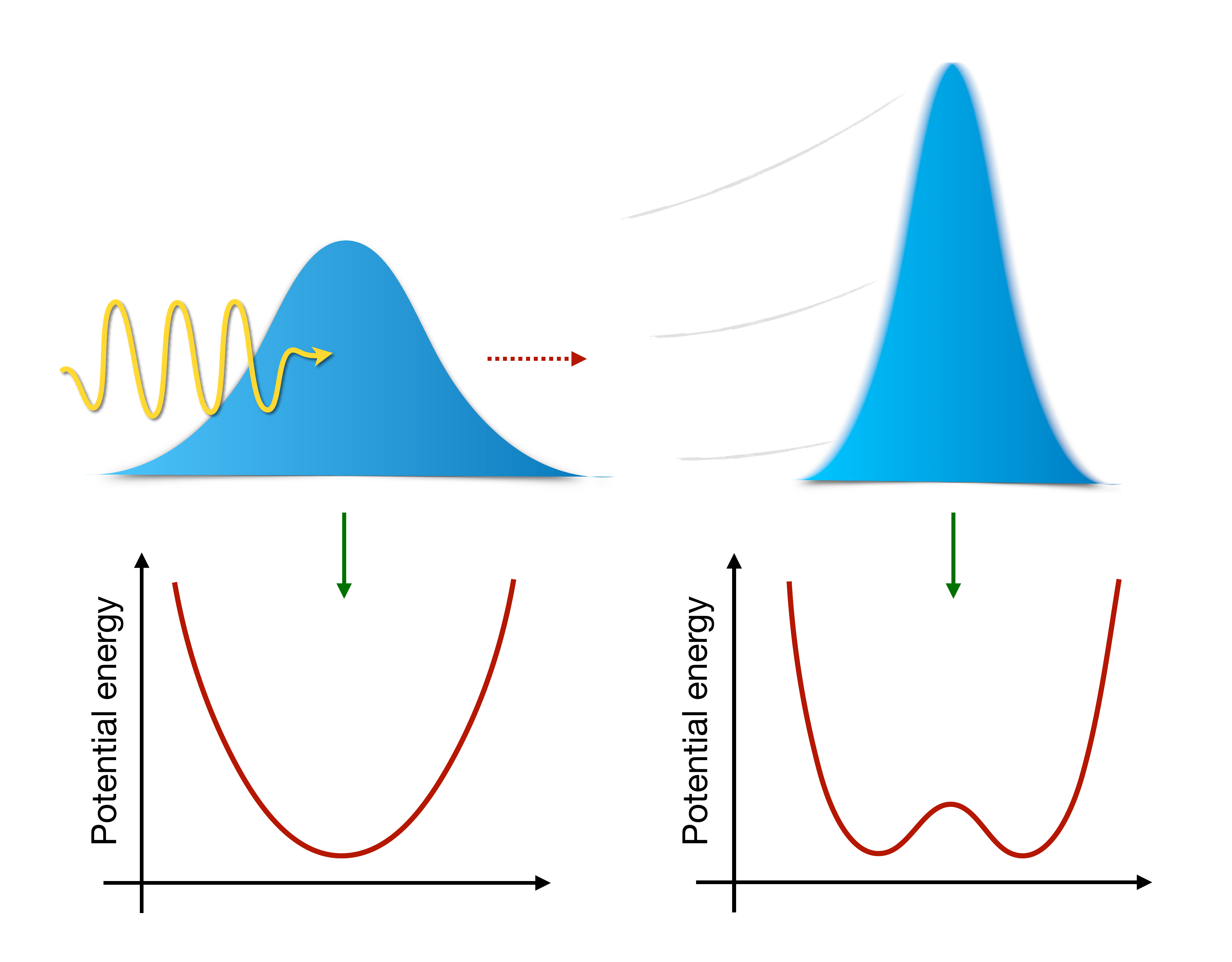}
    \caption{The selective excitation of a phonon mode with a strong laser pulse induces a cooling of its lattice fluctuations, resulting in a narrowing of its density. This narrowing significantly affects the ensemble average of the potential energy $\langle V \rangle$, and can restore the double-well shape of the bare PES.  }
    \label{fi_cooling}
\end{figure}

\section{Phase Transition Induced by Density Cooling}\label{cooling}
\begin{figure*}
    \centering
    \begin{subfigure}{0.48\textwidth}
        \includegraphics[width=\textwidth]{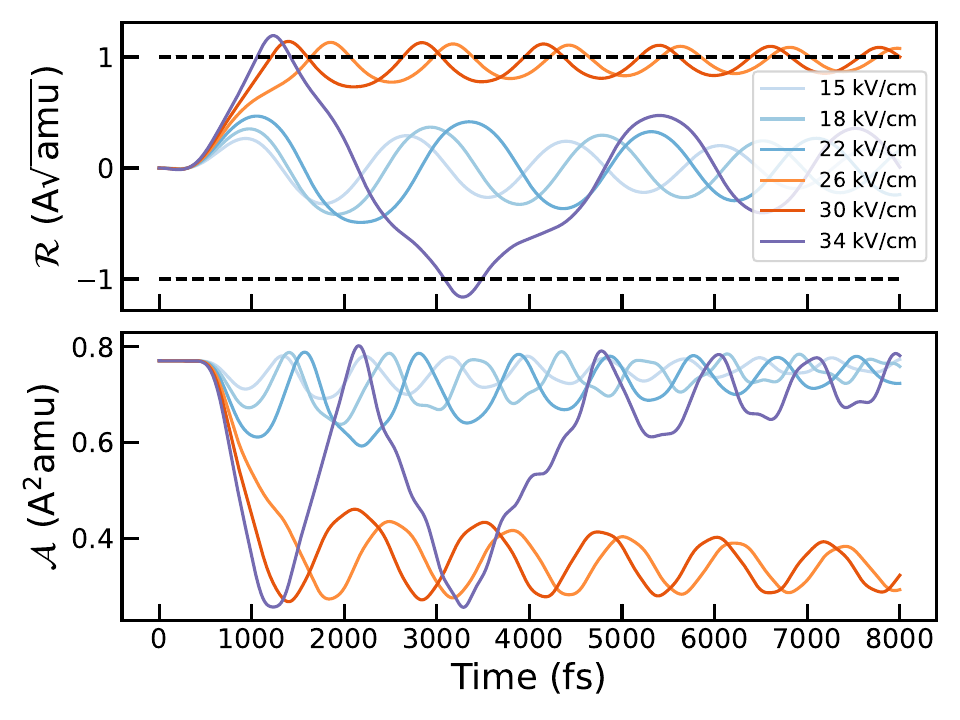}
        \caption{Variable $\ma$}
        \label{}
    \end{subfigure}
    \begin{subfigure}{0.48\textwidth}
        \includegraphics[width=\textwidth]{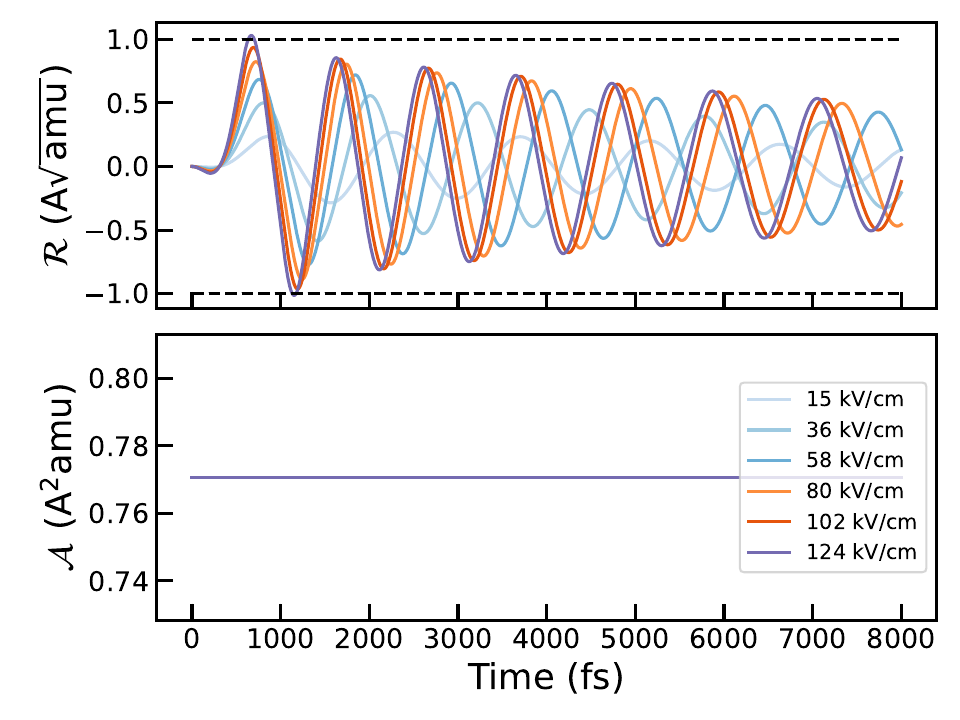}
        \caption{Constant $\ma$}
        \label{A_const}
    \end{subfigure}
    \caption{ The upper and lower panels show the dynamics of $\mr$ and $\ma$, respectively, for different electric field strengths. In the left panels, $\ma$ is evolved according to Eq. \ref{single}, while in the right panels it is kept constant to the initial equilibrium value.  }
    \label{model_transition}
\end{figure*}

In this section, we propose a novel paradigm for nonlinear phononics, which leverages on the cooling of the lattice fluctuations associated to a specific phonon mode to induce a phase transition.
Consider the case of a phonon with a double-well PES. From Eq. \ref{main}, the equation of motion for its lattice fluctuations is given by
\begin{equation}\label{single}
    \dddot{\tilde\ma} + 4 \dot{\tilde{\ma}}\Bigl( \omega^2 + \frac{1}{2}\psi \tilde\ma \Bigr) + \ma\Bigl(2\psi\dot{\tilde\mr}\tilde\mr+ \psi\dot{\tilde\ma} \Bigr) = 0\ ,
\end{equation}
(here, $\chi = 0$ due to symmetry). We aim to study the solution of this equation in the case where a pulse drives the coordinate $\mr$ out of its equilibrium position.
Eq. \ref{single} is a nonlinear third-order differential equation that cannot be solved analytically. An approximate solution can be derived by assuming that $\tilde{\ma}$ is significantly smaller than its equilibrium value, $\ma^{eq}$, an assumption that holds during the initial stages of the dynamics. Under this condition, Eq. \ref{single} simplifies to:
\begin{equation}\label{linear}
    \dddot{\tilde\ma} +  \Omega^2\dot{\tilde{\ma}} + \psi\ma^{eq}\frac{d}{dt}\tilde\mr^2  = 0\ .  
\end{equation}
where $\Omega^2 = 4\omega^2 + \psi \ma^{eq}$ is the frequency of the lattice fluctuations. Integrating Eq. \ref{linear}  and   imposing the initial conditions in Eq. \ref{ic}, we obtain 
\begin{equation}\label{second_order_single_mode}
    \ddot{\tilde\ma} +  \Omega^2\tilde{\ma} + \psi\ma^{eq}\tilde\mr^2  = 0\ .    
\end{equation} 
Since $\mr^2$ is always positive and initially equal to 0, irradiation with a pulse will cause its rapid initial increase.
As a result, $\mathcal{A}$ progressively decreases over time. To illustrate this analytically, consider a Dirac pulse driving the phonon mode. The motion of the mode can be described by $\tilde\mr(t) \sim \theta(t)\mathcal{P}_0t$, where $\mathcal{P}_0$ denotes the finite momentum imparted to the system by the pulse, and $\theta(t)$ is the Heaviside step function.
The solution of Eq. \ref{second_order_single_mode} is given by
\begin{equation}
    \ma(t) = \ma^{eq} - \frac{\psi\ma^{eq}\mathcal{P}_0^2}{\Omega^4}\Bigl( \Omega^2 t^2 + 2\cos(\Omega t)-2 \Bigr)\,
\end{equation}
which denotes a quadratic cooling of the lattice fluctuations. Moreover, the faster the motion of $\tilde\mr$, the more pronounced the cooling effect becomes.
We now illustrate how this mechanism can be exploited to drive a phase transition in materials characterized by the presence of phonon modes with a double-well PES. \\
The ensemble average of the potential energy of a phonon mode in a double well PES is 
\begin{equation}
    \braket{V} = -|\phi| + \frac{1}{2}\psi\mr^2+\frac{1}{2}\psi \ma\ .
\end{equation}
When the barrier height is sufficiently shallow, the initial value of lattice fluctuations, determined through Eq. \ref{sscha}, satisfies $\frac{1}{2}\psi\ma > |\phi|$, leading to the stabilization of the symmetric phase $\mr = 0$. A dynamical cooling of lattice fluctuations brings the system closer to the classical limit, $\ma = 0$, restoring the double-well shape of the potential energy and thereby stabilizing the non-centrosymmetric phase, as illustrated in Fig. \ref{fi_cooling}.\\
The cooling alone does not guarantee an induced phase transition. For a transition to occur, the system must remain self-trapped in a new state characterized by smaller fluctuations, otherwise $\ma$ will grow back toward the equilibrium value. This new state can, for example, correspond to a local minimum of the free energy, which formally represents a solution of the SCHA equations \ref{sscha}, albeit with higher energy than the global symmetric minimum. In such a case, the transition is metastable and persists even after the dynamics have ceased. 
\begin{figure}
    \centering
    \includegraphics[width=\linewidth]{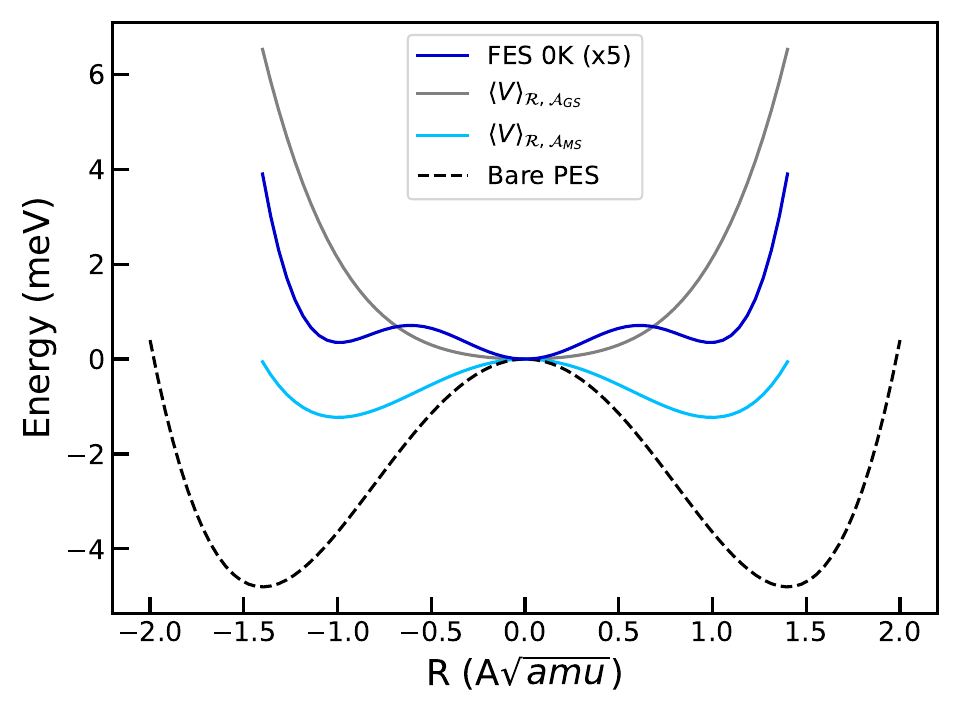}
    \caption{The blue line represents the FES of the one-dimensional model calculated at 0 K, as defined in Eq. \ref{free_surface_eq}. To enhance the visibility of its minima, it has been scaled by a factor of 5. The black dashed line corresponds to the bare potential energy, as defined in Eq. \ref{V_1d}. The grey and light blue lines represent the ensemble-averaged potential energy, computed for $\ma = \ma_{GS}$ and $\ma = \ma_{MS}$, respectively. Notably, the minimum of the grey curve aligns with the equilibrium position of the GS ($\mr_{GS} = 0$), while the minimum of the light blue curve corresponds to the MS ($\mr_{MS} \simeq 1\ \mathrm{\AA\sqrt{amu}}$). }
    \label{fes}
\end{figure}
Alternatively, it could represent a dynamical state with no equilibrium, remaining stable only while the dynamics are ongoing. 
These scenarios were studied in Ref. \cite{libbi2025quantumcoolingabsolutezero} for the specific case of $\mathrm{SrTiO_3}$, but further theoretical and experimental work is needed to determine the conditions that enable self-trapping in a broader class of materials.\\
In the following, we present a numerical example of a phase transition induced by the cooling of lattice fluctuations, based on the solution of a model describing a phonon in the double-well potential given in Eq. \ref{V_1d}.
The phonon’s position evolves according to the first two of Eqs. \ref{tdscha}:
\begin{equation}
    \ddot{\mr} + \gamma\dot{\mr} + \phi\mr + \frac{1}{3!}\psi\mr^3+\frac{1}{2}\psi\mr\ma=\frac{Z^*E}{\varepsilon}\ ,
\end{equation}
while the quantum fluctuations follow Eq. \ref{single}.
In this model, we adopt $V^0= \mathrm{4.8\ meV}$ and $x^0\mathrm{ = 1.4\ \AA\sqrt{amu}}$, which correspond to $\phi=\mathrm{-9.8\ meV\AA^{-2}amu^{-1}}$ and $\psi = 30.0\ \mathrm{meV \AA^{-4}amu^{-2}}$.  The phonon is coupled to the electric field through the effective charge $Z^* = \mathrm{3.0\ amu^{-1/2}}$, and an effective dielectric constant $\varepsilon$ = 1.7. 
\begin{comment}
    3.0 \sim 0.1/np.sqrt(900)
    1.7 = \frac{1+\sqrt{6}}{2}
\end{comment}
To simulate the energy dissipation due to interactions with other phonon modes, the motion is damped with $\gamma=\mathrm{6\cdot10^{-4}\sqrt{meV}\AA^{-1}}$. \\
First, we compute the free energy surface of the model, defined as 
\begin{equation}\label{free_surface_eq}
    F(\mr) = \min_{\Phi}\mathcal{F}(\mr,\Phi)\ .
\end{equation}
It can be readily verified that the minima of the FES are solutions to Eq. \ref{sscha} and therefore represent the equilibrium points of the system.
Fig. \ref{fes} shows the FES computed at 0K (blue line). Unlike the bare PES, defined in Eq. \ref{V_1d} and represented in dashed black line, the FES is made up of three minima. The minimum  at $\mr_{GS}=0$ coincides with the lowest state energy, and is thus the ground state (GS) of the system. The SCHA frequency at this point is $\omega \simeq 0.65$  THz, corresponding to lattice fluctuations $\ma_{GS} = \frac{\hbar(2n+1)}{2\omega} = \mathrm{0.77\ \AA^2amu }$. The minima at $\mr_{MS}\simeq\pm 1\mathrm{\AA\sqrt{uma}}$, instead, have higher energy than the ground state, and represent metastable states (MS), with SCHA frequency of $\sim$1.5 THz and lattice fluctuations $\ma_{MS} = \mathrm{0.34\ \AA^2amu}$. \\
In order to avoid confusion, it is important to clarify that the dynamics of the system follows the ensemble average of the potential energy $\braket{V}$, and not the FES, i.e. $\ddot{\mr} = \braket{\frac{dV}{dR}}\neq \frac{dF}{d\mr}$. This inequality turns into an equality only for the equilibrium points, as shown in Eq. 18 of Ref. \cite{PhysRevB.96.014111}. Specifically, 
\begin{equation}
    \frac{dF}{d\mr}(\mr_{GS}) = \Bigl\langle \frac{dV}{dR} \Bigr\rangle_{\mr_{GS}, \ma_{GS}} = 0\ ,
\end{equation}
and
\begin{equation}
    \frac{dF}{d\mr}(\mr_{MS}) = \Bigl\langle \frac{dV}{dR} \Bigr\rangle_{\mr_{MS}, \ma_{MS}} = 0 \ .
\end{equation}
In other words, these equations indicate that when the value of the lattice fluctuations matches that of a specific minimum of the FES, that minimum will also correspond to a minimum of the ensemble-averaged potential energy. This is clearly illustrated in Fig. \ref{fes}, where the minimum of $\braket{V}_{\mr,\ma_{GS}}$ occurs at 0, while that of $\braket{V}_{\mr,\ma_{MS}}$ is located at $\mathrm{1\ \AA\sqrt{amu}}$.
This insight provides a clearer understanding of the role of lattice fluctuation cooling in a phase transition: it modifies the shape of $\braket{V}$ until its minima coincide with the metastable minima of the free energy, allowing the system to remain in a self-trapped state . \\ 
We will now demonstrate this mechanism through dyanmics simulations on the one-dimensional model. We assume the system starts in thermal equilibrium in the GS and is then perturbed by a Gaussian pulse described by the equation $E(t)= E_0\cos(\omega (t-t_0))\exp^{-\frac{(t-t_0)^2}{2\sigma^2}}$, where the pulse frequency is resonant with the SCHA frequency, $t_0$ = 500 fs and $\sigma$ = 470 fs. \\
Fig. \ref{model_transition} illustrates the temporal evolution of $\mr$ and $\ma$ under varying electric field strengths. For the lowest electric fields (light blue lines), the system exhibits oscillations around the equilibrium position. The dynamics are characterized by an initial cooling of lattice fluctuations, which, however, does not produce any notable impact on the system. At sufficiently high electric field strengths (orange lines), the system undergoes symmetry breaking and begins oscillating around the metastable minimum $\mr_{MS} \simeq 1 , \mathrm{\AA\sqrt{amu}}$. The dynamics of $\ma$ are dominated by a pronounced cooling induced by the pulse irradiation. Notably, $\ma$ does not revert to its initial equilibrium value but instead oscillates around $\ma_{MS} = \mathrm{0.34\ \AA\sqrt{amu}}$. The system is thus self-trapped in the metastable state.
For very large electric fields (violet line), the system undergoes a similar cooling of the lattice fluctuations as seen for intermediate fields; however, its kinetic energy is large enough to escape the local energy minimum. Following a pronounced initial oscillation, the system looses energy, and relaxes towards the GS. \\ 
To highlight the crucial role of lattice fluctuation dynamics in the symmetry-breaking process, we repeated the simulations while keeping their values constant. Figure \ref{A_const} confirms that, under these conditions, the symmetry breaking disappears. This occurs because the system’s potential energy, $\braket{V}$, retains a single-well profile, and the transition to a double-well configuration, as depicted in Fig. \ref{fi_cooling}, does not take place. Importantly, the monotonic behavior of $\braket{V}$ causes the electric field required to achieve $\mr = 1 , \mathrm{\AA\sqrt{amu}}$ to increase by a factor of three.\\
It is useful to represent the phase portrait of the variable \( \ma \) for three different electric field strengths, E = 10, 30,  and  60  kV/cm.
For all field values, the trajectory initially moves toward the lower left, indicating a reduction in lattice fluctuations. At lower field strengths, \( \ma \) oscillates around its equilibrium value and eventually returns to it after a few oscillations. For  E = 30  kV/cm, the phase transition shifts \( \ma \) to a new minimum characterized by reduced fluctuations. At  E = 60  kV/cm, the system exhibits chaotic dynamics, cyclically visiting the symmetry-broken minimum without becoming permanently trapped in it.
\begin{figure*}[t]
    \centering
    \includegraphics[width=\linewidth]{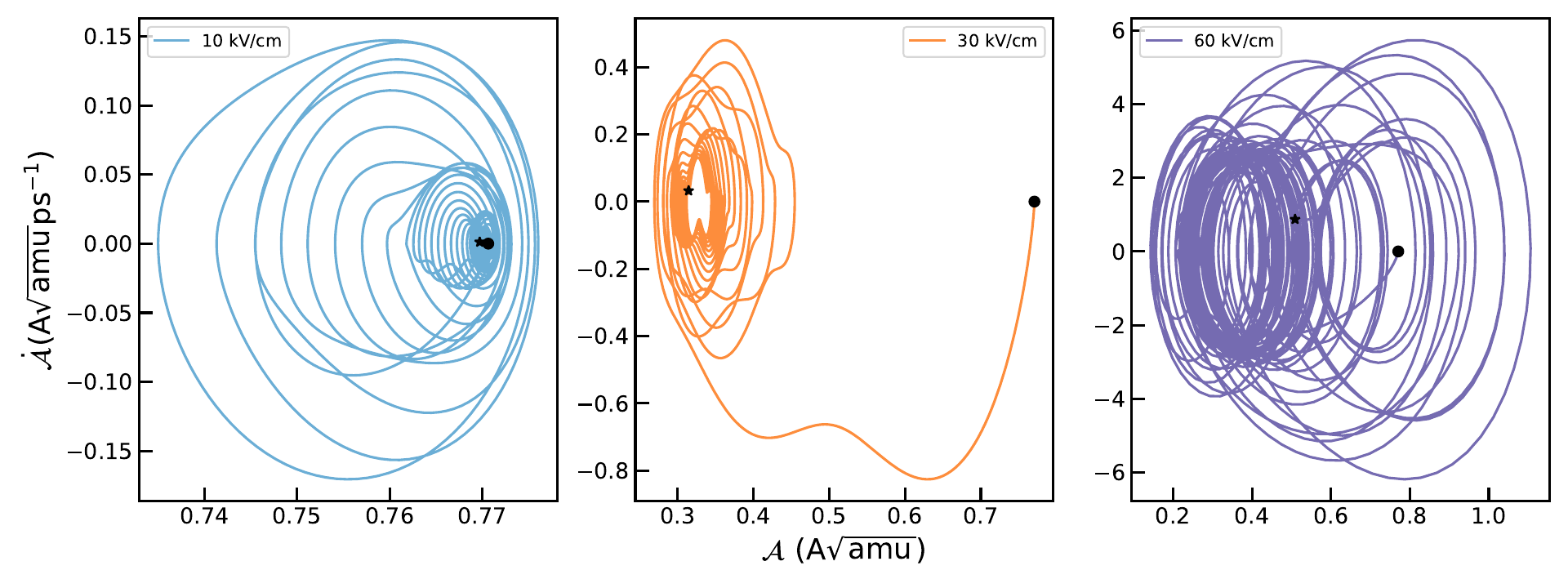}
    \caption{Phase portrait of the dynamics of $\ma$ for different electric fields. The starting point is marked with a circle, while the end point with a star.  }
    \label{attractor}
\end{figure*}

\section{Conclusions}
In this work, we presented a generalized quantum theory of nonlinear phononics, derived by reformulating the TD-SCHA equations through the Taylor expansion of the potential energy up to fourth order and the exact calculation of ensemble averages. This theory not only provides an accurate and efficient framework for simulating quantum nuclear dynamics in materials but also serves as a foundation for deriving minimal models to interpret and predict novel nonlinear phononics mechanisms for controlling material properties.
Building on this new framework, we derived a disentangled equation for the evolution of the diagonal component of lattice fluctuations, which recovers, as a special case, the model proposed in Ref. \cite{fechner_quenched_2024}. 
We utilized this equation to show that a sudden displacement of a phonon mode from equilibrium, induced by a strong pulse, induces the quenching of its lattice fluctuations. Based on this finding, we proposed a new paradigm for nonlinear phononics, leveraging this cooling effect to drive symmetry breaking in materials characterized by phonons with a double-well potential energy surface.

\section{Acknowledgements}
We acknowledge useful discussions with Prof. Lorenzo Monacelli. This research was funded in part by the Swiss National Science Foundation (SNSF, Mobility fellowship P500PT\_217861), the National Science Foundation Harnessing the Data Revolution Big Idea under Grant No. 2118201, and the Harvard SEAS Dean’s Competitive Fund for Promising Scholarship. 
Computational resources were provided by the FAS Division of Science Research Computing Group at Harvard University.
\appendix

\section{Mass rescaling}\label{scaling_rules}
In this work, we follow the mass-scaling convention proposed in Ref. \cite{PhysRevB.107.174307}. Table \ref{scaling} summarizes the scaling rules for all variables defined in the main text. As a general guideline, all lengths are multiplied by $\sqrt{m}$, and momenta are divided by $\sqrt{m}$. Consequently, any quantity representing a derivative with respect to position is scaled by $m^{-\frac{n}{2}}$, where $n$ is the order of the derivative.
\begin{table}[h!]
    \centering
\renewcommand{\arraystretch}{1.7} 
    \setlength{\tabcolsep}{12pt} 
    \begin{tabular}{|c|}
        \hline
         Scaling rules \\
         \hline
         $\mr_i = \sqrt{m_i}\bar{\mr}_i$  \\
         $\mathcal{P}_i = \frac{\bar{\mathcal{P}}_i}{\sqrt{m_i}}$ \\
         $\ma_{ij} = \sqrt{m_i}\bar{\ma}_{ij}\sqrt{m_j}$\\
         $\mathcal{B}_{ij} = \frac{1}{\sqrt{m_i}} \bar{\mathcal{B}}_{ij} \frac{1}{\sqrt{m_j}}$ \\
         $\Gamma_{ij} = \bar{\Gamma}_{ij}$ \\
         $\phi_{ij} = \frac{\bar{\phi}_{ij}}{\sqrt{m_im_j}}$ \\
         $\chi_{ij} = \frac{\bar{\chi}_{ij}}{\sqrt{m_im_jm_k}}$ \\
         $\psi_{ij} = \frac{\bar{\psi}_{ij}}{\sqrt{m_im_jm_km_l}}$ \\
         $\braket{f}_i = \frac{\braket{\bar{f}_i}}{\sqrt{m_i}}$ \\
         $\kappa_{ij} = \frac{\bar{\kappa}}{\sqrt{m_im_j}}$ \\ \hline
    \end{tabular}
    \caption{Scaling rules applied throughout this work. Variables marked with a bar are those not scaled by masses. }
    \label{scaling}
\end{table}

\section{Equation for $\mathbf{A_{AFD}}$}\label{eq_A}

In this section, we provide the complete derivation of the model described in Eqs. \ref{main}, \ref{kmu1}, and \ref{kmu_dot1}. We begin by rewriting Eqs. \ref{tdscha} in the phonon basis, utilizing the projection rule discussed in the main text, which we restate here for reference:
\begin{equation}
    T_{\mu\nu...\sigma} = \sum_{ij... k}T_{ij...k}\ e_{\mu i}e_{\nu j}...e_{\sigma k}
\end{equation}
Substituting this expression into Eq. \ref{tdscha} yields equivalent equations where the Greek indices replace the Cartesian ones. For clarity, we rewrite them below:
\begin{equation}\label{tdscha_mu}
    \begin{cases}
        \dot{R}_\mu = P_{\mu} \\
        \dot{P_{\mu}} = \braket{f}_{\mu} \\
        \dot{\ma}_{\mu\nu} = \Gamma_{\mu\nu} + \Gamma_{\nu\mu} \\
        \dot{\mathcal{B}}_{\mu\nu} = -\sum_\tau\braket{\partial^2V}_{\mu\tau}\Gamma_{\tau\nu} - \sum_\tau\braket{\partial^2V}_{\nu\tau}\Gamma_{\tau\mu}\\
        \dot{\Gamma}_{\mu\nu} = \mathcal{B}_{\mu\nu} - \ma_{\mu\tau}\braket{\partial^2V}_{\tau\nu}
    \end{cases}\ .
\end{equation}
The diagonal components of $\ma$ and $\mathcal{B}$ evolve as follows: 
\begin{equation}\label{A_mu_dot}
    \dot{\ma}_{\mu\mu} = 2\Gamma_{\mu\mu}
\end{equation}
\begin{equation}
    \dot{\mathcal{B}}_{\mu\mu} = -2\sum_{\tau}\kappa_{\mu\tau}\Gamma_{\tau\mu} \ .
\end{equation}
Neglecting the off-diagonal couplings between the curvature and lattice fluctuations, as expressed in Eq. \ref{approx_gamma_text}, we find:
\begin{equation}\label{b_mu_dot}
    \dot{\mathcal{B}}_{\mu\mu} \simeq 2\kappa_{\mu\mu}\Gamma_{\mu\mu} = -\kappa_{\mu\mu}\dot{\ma}_{\mu\mu}\ .
\end{equation}
By differentiating Eq. \ref{A_mu_dot} twice and combining it with the last equation in Eqs. \ref{tdscha_mu} as well as Eq. \ref{b_mu_dot}, we obtain:
\begin{equation}
    \dddot{\ma}_{\mu\mu} + 2\dot{\ma}_{\mu\mu}\kappa_{\mu\mu} + 2\sum_{\tau}\dot{\ma}_{\mu\tau}\kappa_{\tau\mu} + 2\sum_{\tau}\ma_{\mu\tau}\dot{\kappa}_{\tau\mu} = 0 \ .
\end{equation}
This simplifies to: 
\begin{equation}\label{Amumu}
    \dddot{\ma}_{\mu\mu} + 4\dot{\ma}_{\mu\mu}\kappa_{\mu\mu} + 2\ma_{\mu\mu}\dot{\kappa}_{\mu\mu} = 0 \ ,
\end{equation}
when Eq. \ref{approx_a_text} is imposed.
Next, we express Eq. \ref{curv} in phonon coordinates and differentiate it:
\begin{equation}\label{kmu}
    \kappa_{\mu\mu} = \phi_{\mu\mu} + \sum_\tau\chi_{\mu\mu\tau}\mr_{\tau} + \sum_{\tau\sigma}\frac{1}{2}\psi_{\mu\mu\tau\sigma}(\mr_{\tau}\mr_{\sigma} + \ma_{\tau\sigma} ) \ ,
\end{equation}
\begin{equation}\label{kmu_dot}
    \dot{\kappa}_{\mu\mu} = \sum_\tau\chi_{\mu\mu\tau}\dot{\mr}_{\tau} + \sum_{\tau\sigma}\frac{1}{2}\psi_{\mu\mu\tau\sigma}(2\dot{\mr}_{\tau}\mr_{\sigma} +  \dot{\ma}_{\tau\sigma} ) \ .
\end{equation}
Since Eqs. \ref{kmu} and \ref{kmu_dot} are second-degree polynomials, they can be systematically re-expressed as a Taylor series expansion around the SSCHA equilibrium configuration without any loss of generality. Defining the displacement from the equilibrium position as $\tilde{\mr}_{\mu} = \mr_{\mu} - \mr_{\mu}^{eq}$ and the deviation in lattice fluctuations relative to the equilibrium as $\tilde{\ma}_{\mu\nu} = \ma_{\mu\nu} - \ma_{\mu\nu}^{eq}$, we can reformulate the curvature and its time derivative as reported in Eqs. \ref{kmu1} and \ref{kmu_dot1} of the main text. We have used the relation:
\begin{equation}
\omega^2_{\mu} = \Bigl\langle \frac{\partial^2 V}{\partial \mr_{\mu}^2} \Bigr\rangle \Big|_{\mr^{eq}, A^{eq}} \ ,
\end{equation}
which stems directly from the second Eqs. \ref{sscha}.
As mentioned in the main text, an external potential $V_{ext}$, which is linear in the electric field, does not affect the ensemble average of the curvature and, therefore, does not directly couple with the lattice fluctuation matrix. Consequently, the lattice fluctuations can only be influenced indirectly by driving the displacement of the centroids, as shown in Eqs. \ref{main}–\ref{kmu_dot1}. 
By noting that $\dot{\ma}=\dot{\tilde\ma}$ and $\ddot{\ma}=\ddot{\tilde\ma}$, it is possible to turn Eq. \ref{Amumu} into Eq. \ref{main}.
\\

\bibliographystyle{ieeetr}

\end{document}